\documentclass{ws-ijmpe}
\usepackage[super,compress]{cite}
\begin{document}

\markboth{Richard Herrmann}{Curvature Interaction in Collective Space}

\catchline{}{}{}{}{}

\title{Curvature Interaction in Collective Space}
\author{\footnotesize Richard Herrmann}
\address{GigaHedron, Berliner Ring 80, D-63303 Dreieich\\
herrmann@gigahedron.com}

\maketitle

\begin{history}
\received{Day Month Year}
\revised{Day Month Year}
\end{history}

\begin{abstract}
For the Riemannian space, built from the collective coordinates used within nuclear models, an additional interaction with the metric is investigated, using the collective equivalent to Einstein's curvature scalar. The coupling strength is determined using a fit with the  AME2003 ground state masses. An extended finite-range droplet model inclu\-ding curvature is introduced, which generates significant improvements for light nuclei and nuclei in the trans-fermium region. 
\end{abstract}


\ccode{PACS numbers:21.60Ev;02.40.Ky;21.10.Dr; 21.10.Gv; 04.62.+v}


\section{Introduction}
The use of collective models for a description of collective aspects of nuclear motion 
has proven considerably successful during the past decades. 
Calculating life-times of heavy nuclei \cite{mye66}, \cite{gru}, \cite{sta12},
fission yields \cite{lus},  giving insight into phenomena like cluster-radioactivity \cite{san}, \cite{poe}, bimodal fission or modeling the ground state properties of triaxial nuclei \cite{moe08} -
remarkable results have been achieved by introducing an appropriate
set of collective coordinates, like length, deformation, neck or
mass-asymmetry \cite{mar} for a given nuclear shape and investigating its dynamic properties.

We start with the classical Hamiltonian function
\begin{equation}
\label{h0h0}
H = T+V_0
\end{equation} 
introducing a collective potential $V_0$, depending on $N$ collective coordinates $\{q^i, i=1,...,N\}$, 
\begin{equation}
V_0(q^i) = E_{\textrm{macro}}(q^i) + E_{\textrm{mic}}(q^i)
\end{equation} 
with a macroscopic contribution $E_{\textrm{macro}}$ based on e.g. the liquid drop model and a microscopic contribution $E_{\textrm{mic}}$, which mainly contains the shell and pairing energy,
and the classical kinetic energy $T$
\begin{equation}
T = \frac{1}{2} B_{ij}\, \dot{q}^i \dot{q}^j
\end{equation} 
with collective mass parameters $B_{ij}$.

There are several common methods 
to generate the collective mass parameters $B_{ij}$, e.g.  the cranking model \cite{ing} or irrotational flow models are used.  

We want to emphasize the fact, that via the relation 
\begin{equation}
B_{ij} = m_A \, g_{ij}= m_u  A \,g_{ij} \label{bijgij}
\end{equation} 
($m_A$ is the mass of the nucleus, $m_u=931.5$ MeV is the mass unit  and $A$ is the number of nucleons)  
the collective masses may be interpreted geometrically, 
defining  the metric tensor $g_{ij}$, which fully determines the geometric properties of the collective Riemannian space.

Quantization of the classical Hamiltonian\cite{pod} 
results in the collective Schr\"odinger equation 
\begin{eqnarray}
       \hat{S}_0 \Psi(q^i,t) =\Big(-\frac{\hbar^{2}}{2 \, m_A} \frac{1}{\sqrt{g}} \partial_{i}
        g^{ij} \sqrt{g} \partial_{j}  -i\hbar  \partial_{t} + V_0\Big)\Psi(q^i,t)  = 0
\end{eqnarray}
which is the central starting point for a discussion of nuclear collective phenomena.

An alternative approach starts with the Lagrangian density
${\cal L}_0$
\begin{equation}
{\cal L}_0 = \frac{\hbar^2}{2 \, m_A} g^{ij}(\partial_{i} \Psi^*)(\partial_{j} \Psi) 
    +\frac{i \hbar}{2 }(\Psi^* \frac{\partial\Psi}{\partial t} - \frac{\partial\Psi^*}{\partial t}\Psi) 
- \Psi^* V_0 \Psi
\end{equation} 
Variation with respect to $\Psi^*$ and $\Psi$ yields the above Schr\"odinger equation.

From this point of view it is remarkable, that  obvious extensions of this Lagrangian density have been discussed in other branches of physics, e.g. cosmology or string theory, but have been neglected within the framework of nuclear collective models until now.

Therefore in the following we will introduce a curvature interaction in collective space and discuss the consequences for the prediction of nuclear binding energies.

\section{Curvature in collective space}
In order to investigate the influence of non vanishing curvature in collective space, we  consider an additional interaction with the collective metric, which is determined by the collective mass parameters. 

We extend the Lagrangian density 
\begin{equation}
{\cal L} = {\cal L}_0 - \frac{\hbar^2}{2 \, m_A} \xi \Psi^* R \Psi
\end{equation} 
introducing the Einstein curvature scalar $R$ as an invariant measure for collective curvature.
The coupling strength is parametrized with $\xi$.

Variation of this Lagrangian density results in  an additional potential term 
\begin{equation}
V = V_0 +  \frac{\hbar^2}{2 m_A} \xi  R 
\end{equation} 
Since the collective mass parameters are known, $R$ can be calculated.    
As a starting point the Riemann curvature tensor  is given by
\begin{equation}
{R^\alpha}_{\eta \beta \gamma} = 
  \partial_{\gamma}\Gamma^\alpha_{\beta\eta}
 -\partial_{\beta}\Gamma^\alpha_{\eta\gamma}
+ \Gamma^\alpha_{\tau\gamma}\Gamma^\tau_{\beta\eta}
- \Gamma^\alpha_{\tau\beta}\Gamma^\tau_{\gamma\eta}
\end{equation} 
with the Christoffel symbols of second kind \cite{adl}
\begin{equation}
\Gamma^\mu_{\kappa \sigma} = 
\frac{1}{2}g^{\nu\mu}(\partial_{\kappa}g_{\nu\sigma} + \partial_{\sigma}g_{\nu\kappa}-\partial_{\nu}g_{\kappa\sigma})
\end{equation} 
The Riemann curvature tensor
may be contracted to get the Ricci tensor
\begin{equation}
R_{\eta \gamma} = {R^\alpha}_{\eta \alpha \gamma}  
\end{equation} 
and finally we obtain the Einstein curvature scalar $R$  via:
\begin{equation}
R = {R^\eta}_\eta 
\end{equation} 
The explicit  form of this curvature term 
depends on the specific choice of collective coordinates.  

In order to examine the
consequences and physical interpretation of this additional new term we 
choose the symmetric two-center shell model including elliptical deformations \cite{sch}, which can be solved fully analytically. This model  is widely used in the description of symmetric fusion reactions
and contains the Nilsson model as a limiting case, which will turn out to be a useful property for a physical interpretation.
\section{Exact solution for the symmetric two-center shell model}
\begin{figure}
\begin{center}
\includegraphics[width=88mm]{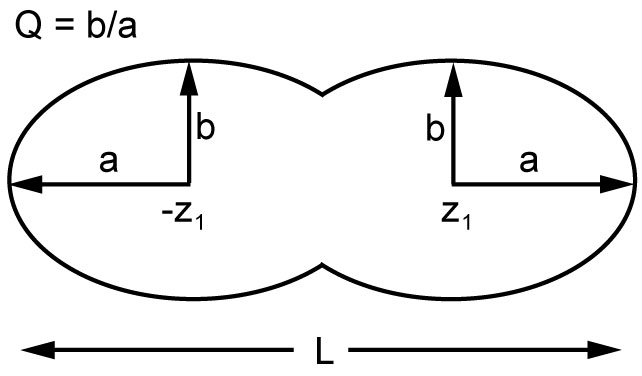}\\
\caption{\label{fig1} geometry of the symmetric two-center shell model}
\end{center}
\end{figure}
As an illustrative, exactly solvable scenario we consider the nuclear shape 
given by two intersecting rotationally symmetric ellipsoids. Introducing 
two collective coordinates $q^i$ namely, 
the ellipsoidal deformation $Q=b/a$ and the 
total elongation $L$ the shape $ P(z,q^i)$  is given by (see figure {\ref{fig1}}):
\begin{equation}
 P(z,q^i) = Q \sqrt{a^2 - (z \mp z_1)^2}
\end{equation}
where the  geometric quantities semi axis $a$ and center position of ellipsoids $z_1$ are determined by the 
definition of $L$ and by the requirement of
volume conservation, which yields in the case of connected fragments  
with $R_0 = r_0 A^{1/3}$:
\begin{eqnarray}
 L &=& 2(a+z_1)\\
 \textrm{volume} &=& (4/3) \pi {R_0}^3 = 2 \,(2/3) \, \pi Q^2 (a^3 + \frac{3}{2} a^2 z_1 - \frac{1}{2} z_1^3) \label{vol}
\end{eqnarray}
These equations may be simplified introducing the dimensionless quantities  
\begin{eqnarray}
\label{aaa}
 \alpha &=& Q^{2/3} \, a / R_0 \\
 \gamma_1 &=& Q^{2/3} \, z_1/R_0  \\
 \lambda &=&  Q^{2/3} \, L/(2 R_0)  \label{lam}
\end{eqnarray}
Equations (\ref{aaa})-(\ref{lam}) define a transformation to a new set of coordinates $(\lambda,Q)$. 
\\
We obtain:
\begin{eqnarray}
 \alpha &=& \frac{2 + \lambda^3}{3\,\lambda^2}\\
 \gamma_1 &=& \frac{2\,({\lambda }^3  -1  ) }
  {3\,{\lambda }^2}
\end{eqnarray}
which is independent of $Q$. Thus, the shape geometry is fully determined for a given set of collective coordinates $(\lambda,Q)$.

$1 \leq \lambda \leq 4^{1/3}$ describes connected fragments, where $\lambda=1$ is the compound nucleus and $\lambda = 4^{1/3}$ is the scission point and 
$\lambda> 4^{1/3}$ describes separated fragments. $Q < 1$ describes prolate and $Q>1$ oblate shapes.

We now apply the Werner-Wheeler-formalism\cite{wwm} 
to calculate the collective masses $B_{ij}$, which are directly correlated to the metric tensor $g_{ij}$ according to (\ref{bijgij}). We choose this method, since masses are determined by shape geometry only
and the procedure itself is well defined.
Using the abbreviation
\begin{equation}
\Theta = \ln (\frac{4 - {\lambda }^3}{4 + 2\,{\lambda }^3})  
\end{equation}
the components of  the metric tensor $g_{ij}$ result as
\begin{eqnarray}
g_{\lambda \lambda} &=& \frac{{R_0}^2{( 2 + {\lambda }^3 ) }^2} 
{324\,Q^{4/3}\,{\lambda }^{12}}  
{\times}  \\
& & \Big( 3\,(  Q^2 -16 ) \,{\lambda }^9   
    +  4\,( Q^2 -4) \,
   [ 24\,{\lambda }^3 - 3\,{\lambda }^6 + {( {\lambda }^3 -4 ) }^2\,( 2 + {\lambda }^3 ) \,
      \Theta ] \Big) \nonumber \\
 g_{\lambda Q} &=& -\frac{{R_0}^2}{108\,Q^{7/3}\,{\lambda }^2}  
( 2 + \lambda^3 ) \, \Big(6\,\lambda^3 + Q^2\,( 4 - {\lambda }^3 )\Big)   \\
 g_{Q Q} &=&
\frac{{R_0}^2}{810\,Q^{10/3}\,\lambda }
    \Big( 12\,{\lambda }^3\,
       ( 5 + {\lambda }^3 )  + 
      Q^2\,( 40 - 5\,{\lambda }^3 + 
         {\lambda }^6 )  \Big)  
\end{eqnarray}
A coordinate transformation from the coordinate set $(\lambda, Q)$ to the original $(L,Q)$ using
\begin{equation}
g_{ij} = \frac{\partial x^m}{\partial x^i}\frac{\partial x^n}{\partial x^j}g_{mn} \label{trans}
\end{equation}
yields the final result for connected shapes in the range $1 \le \lambda \le 4^{1/3}$
\begin{eqnarray}
 g_{LL}&=&\frac{{( 2 + {\lambda }^3 ) }^2}{1296{\lambda }^{12}} 
 \, \times 
  \\
& & 
 \Big( 3(Q^2-16) \lambda^9 + 
      4 (Q^2-4) [ 8 \lambda^3 ( 3-\lambda^3) + 
         ( 4-\lambda^3)^2 ( 2 + \lambda^3)  \Theta ] \Big)  \nonumber\\ 
 g_{LQ}&=&\frac{R_0\,( 2 + {\lambda }^3 )}{1944\,Q^{5/3}{\lambda }^{11}} 
 \, \times 
 \\
& & 
\Big( 3{\lambda }^3[ (Q^2+14) {\lambda }^9 
       + 8(Q^2 -4) ( 16 + 6\,{\lambda }^3 - 3{\lambda }^6 )  ]  +
 \nonumber \\
& & 
        8(Q^2 -4) ( -8 - 2{\lambda }^3 + {\lambda }^6 )^2 \Theta \Big) 
\nonumber\\
 g_{QQ}&=&\frac{{R_0}^2}{7290\,Q^{10/3}{\lambda }^{10}} 
 \, \times
 \\
& & 
 \Big( ( 48 + 69\, Q^2 ) {\lambda }^{15} + 
      ( Q^2 -4) [ 15{\lambda }^3 ( 256 + 224{\lambda }^3 - 31{\lambda }^9 )  + 
 \nonumber \\
& & 
         40{( {\lambda }^3 -4 ) }^2{( 2 + {\lambda }^3 ) }^3 \Theta ] \Big)
\quad\quad\quad\quad\quad\quad\quad\quad\quad\quad 1 \leq \lambda \leq 4^{1/3}
\nonumber 
\end{eqnarray}
A similar calculation for separated fragments with $\lambda \ge 4^{1/3}$ yields
\begin{eqnarray}
\label{gg0}
g_{L L} &=& \frac{1}{4} \\
g_{L Q} &=& \frac{R_0/3}
  {2^{1/3}\,Q^{5/3}}\\
\label{gg2}
g_{Q Q} &=&
\frac{2^{1/3}\,(12 + Q^2){R_0}^2} {45\,Q^{10/3}}
\quad\quad\quad\quad\quad\quad\quad\quad\quad\quad\quad   \lambda > 4^{1/3}
\end{eqnarray}
\begin{figure}
\begin{center}
\includegraphics[width=126mm]{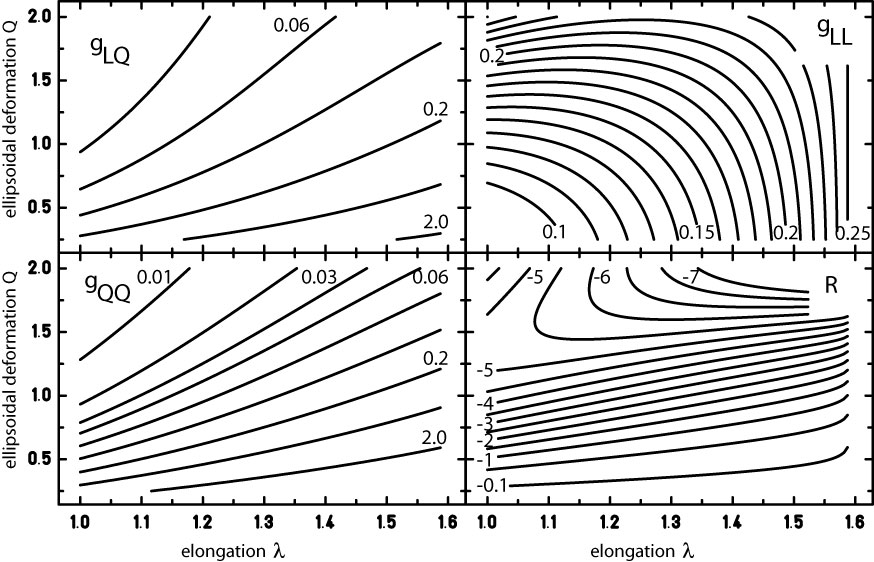}\\
\caption{\label{fig2} metric tensor components 
$g_{LQ}$ (upper left), $g_{LL}$ (upper right), $g_{QQ}$ (lower left) and curvature 
scalar $R$ (lower right) for the symmetric two center shell model, normalized setting $R_0 = 1$}
\end{center}
\end{figure}
Given the metric tensor $g_{ij}$, the curvature scalar $R(L,Q)$ can easily be calculated. 

Figure \ref{fig2} shows the elements of the metric tensor $g_{ij}$ and the resulting curvature 
scalar $R$. 
For connected fragments with $\lambda \le 4^{1/3}$ a non vanishing curvature scalar $R(L,Q)$ exists. 
This is a direct consequence of the volume conservation condition (see (\ref{vol})). 

For prolate and moderately oblate shapes ($Q\le 1.7$) with a fixed Q,  the curvature scalar starts 
with a negative value, which tends to $0$  with increasing $\lambda$ up to the scission point.
For oblate shapes ($Q\ge 1.7$) with  a fixed Q, $R$ decreases with increasing $\lambda$ down to the scission point.

For separated fragments we obtain $R=0$. This discontinuity of the curvature scalar at the scission point is a direct consequence of the underlying
simple geometry, the derivative of the shape is not defined at the contact point of the two ellipsoids. 
This can be avoided by smoothing the shape appropriately, resulting in a smooth curvature term at the scission point,  the resulting model has to be solved numerically though.

The curvature minimizing shape is not a sphere, but a slightly deformed, oblate shape, due to the fact, that the subject of our considerations is not the curvature of a given shape, but curvature of the collective space, generated by the Werner-Wheeler-masses. 
\\
In case of  a single deformed ellipsoid ($\lambda=1$) $R$ is explicitly given as:
\begin{equation}
 R(Q) = 
\frac{-720\,Q^{16/3}\,( -25 + 36\,\ln (2) ) \,( -67 + 96\,\ln (2) ) }
  {R_0^2\,(2 + Q^2)^2 \, [ -2 + (Q^2-4) \, ( -67 + 96\,\ln (2) )  ] ^2}
\end{equation}
and finally, for a sphere ($Q=1$)  this reduces to $R_{\textrm{sphere}}$:
\begin{equation}
 R_{\textrm{sphere}} = -4.3599 \frac{1}{R_0^2}
\end{equation}

Since $R$ is an invariant  under coordinate transformations, the shape 
may be described by any appropriate set of coordinates, which obey the transformation rule given in (\ref{trans}). Consequently, for 
the symmetric two center shell model, which we discussed here as an example, the coordinate sets $(L,Q)$,$(\lambda,Q)$, $(\lambda,\beta = 1/Q)$ or $(\Delta z, \beta)$ where $\Delta z$ is 
the two-center distance, are equivalent. They lead to 
different mass parameters, but yield the same $R$. In that sense, the curvature scalar is a unique, outstanding
property of a given shape geometry.   
\section{Determination of the coupling constant}
In order to get an estimate for the curvature coupling constant, we will now 
investigate the influence of an additional curvature term by a fit of 
experimentally known ground state masses of nuclei.\\ \\
For reasons of simplicity, we assume the ground state of nuclei being of ellipsoidal form only,
neglecting higher order multipoles. Therefore the shapes are described by $(\lambda=1,Q)$, depending on only one
 collective coordinate $Q$.

We define the relative curvature energy $B_R$ with respect to the spherical compound nucleus $(\lambda=1,Q=1)$ as 
\begin{eqnarray}
B_R&=& R(\lambda=1,Q)/R_{\textrm{sphere}}\\
       &=& 9\, Q^{16/3}\Big( \frac{199-288 \ln(2)}{(2+Q^2)[266-67 Q^2 + 96 (Q^2-4) \ln(2)]}\Big)^2
\end{eqnarray} 
An additional curvature potential term $V_R$ is defined
\begin{eqnarray}
 V_R(a_R, Q) &=& +\frac{\hbar^2}{2 m_A} \xi R_{\textrm{sphere}}B_R\\
        &=& -a_R B_R A^{-5/3} 
\end{eqnarray} 
where we introduced the curvature-energy constant $a_R$, which will be determined now.
\begin{table}
\caption{\label{frldm}determination of volume-energy $a_v$, volume-asymmetry $k_v$, surface-energy $a_s$, 
surface-asymmetry $k_s$ and curvature energy $a_R$ constants within the original FRLDM, FRLDM2003 fitted with AME2003 
experimental masses and FRLDMC, which corresponds to FRLDM plus curvature term
and resulting root-mean-square deviations $\Delta_{\textrm{rms}}$
from AME2003 experimental data   
}
\begin{tabular}{r|rrr}
constants&FRLDM&FRLDM2003&FRLDMC\\
\hline\noalign{\smallskip}
$a_v$&16.00126 MeV&16.00496 MeV& 16.01890 MeV\\
$k_v$& 1.92240 MeV& 1.93167 MeV&  1.92882 MeV\\
$a_s$&21.18466 MeV&21.18770 MeV& 21.25974 MeV\\
$k_s$& 2.345   MeV& 2.35968 MeV&  2.34955 MeV\\
$a_R$& -          & -         &529.95850 MeV\\
$\Delta_{\textrm{rms}}$& 0.821 MeV    & 0.815 MeV   &  0.764 MeV\\
\end{tabular}
\end{table}
Our choice for an appropriate macroscopic model is the finite range liquid drop model FRLDM. It is widely used and documented in detail \cite{nix}.

We shall vary only a subset of parameters, namely, the volume-energy $a_v$, the volume-asymmetry $k_v$, 
the surface-energy $a_s$ and the surface-asymmetry $k_s$ constants, which generate the major contributions for the  
calculated masses, keeping all other parameters at their original values.

We define the finite range liquid drop model with curvature (FRLDMC):
\begin{eqnarray}
 \textrm{FRLDMC}(a_v, k_v, a_s,k_s, a_R,Q) &= \textrm{FRLDM}(a_v, k_v, a_s,k_s,Q)+ V_R(a_R,Q)
\end{eqnarray} 
Theoretical masses $m_{\textrm{th}}$ are then obtained, including the microscopic  corrections $E_{\textrm{mic}}$
\begin{eqnarray}
 m_{\textrm{th}} &= \textrm{FRLDMC} + E_{\textrm{mic}}
\end{eqnarray} 
and are compared with the AME2003 experimental masses \cite{aud}.
For conversion from quadrupole moments $\beta_2$ to 
ellipsoidal deformations $Q$ we use the relation:   
\begin{equation}
Q = 1 - \frac{3}{2}\sqrt{\frac{5}{4 \pi}}\beta_2
\end{equation} 
As a measure for the quality of the fit, we tabulate the root mean square deviation
\begin{equation}
\Delta_{\textrm{rms}} =    \sqrt{\frac{1}{N}\sum^N (m_{\textrm{exp}}-m_{\textrm{th}})^2}
\end{equation} 
Results are collected in table \ref{frldm}. The first column lists the original FRLDM parameter set, followed
by results for FRLDM2003, which corresponds to an actualized FRLDM-parameter set for AME2003 masses 
and finally results for FRLDMC, which corresponds to the original FRLDM including the curvature term are presented. 

The corresponding $\Delta_{\textrm{rms}} $-values indicate a significant improvement of the new, extended FRLDMC-model. Especially for light nuclei and
in the region of trans lead elements improvements are significant, as shown in table 3, where errors
for different $Z$-regions are listed.

For light nuclei this is due to the $A^{-5/3}$ behavior of the curvature energy, since this term 
contributes most to the total binding energy for light nuclei, 
e.g. for $^{16}\textrm{O} = 5.21 $ MeV, while for heavy nuclei, this term becomes negligible e.g. for $^{208}\textrm{Pb} = 0.07 $ MeV.
The improvement  indicates, that the additional Riemann curvature term is a useful extension for a collective model.  

Since overestimating masses for heavy nuclei is a known shortcoming of FRLDM, Nix et al.\cite{nix} introduced the finite 
range droplet model (FRDM), whose major improvement is an additional empirical exponential term of the form
\begin{equation}
 -C A \exp^{-\gamma A^{1/3}} \bar{\epsilon} 
\end{equation}
Using original parameters, this term simulates an A-dependence, which is close to the 
collective curvature term, derived in this work.
\begin{table}
\caption{\label{frdm}determination of volume-energy $a_v$, surface-energy $a_s$, 
charge-asymmetry $c_a$ and curvature energy $a_R$ constants within the original FRDM, FRDM2003 fitted with AME2003 
experimental masses and FRDMC, which corresponds to FRDM plus curvature term
and resulting root-mean-square deviations $\Delta_{\textrm{rms}}$
from AME2003 experimental data   
}
\begin{tabular}{r|rrr}
constants&FRDM&FRDM2003&FRDMC\\
\hline\noalign{\smallskip}
$a_v$&16.247 MeV&16.2401 MeV& 16.2467 MeV\\
$a_s$&22.92  MeV&22.8812 MeV& 22.9159 MeV\\
$c_a$& 0.436 MeV& 0.4368 MeV&  0.4332 MeV\\
$a_R$& -          & -       &172.676 MeV\\
$\Delta_{\textrm{rms}}$& 0.679 MeV    & 0.674 MeV   &  0.655 MeV
\end{tabular}
\end{table}
Therefore we expect a reduced influence of an additional curvature term within the framework of FRDM.
\\
 To
proof this hypothesis, 
we define the finite range drop model with curvature (FRDMC) and vary with respect to the subset of most important parameters,
$a_v$ volume-energy, $a_s$ surface-energy and $c_a$ charge-asymmetry constants, keeping all other parameters 
fixed at their original values.
\begin{equation}
 \textrm{FRDMC}(a_v, a_s,c_a,a_R,Q) = \textrm{FRDM}(a_v,a_s,c_a,Q)+ V_R(a_R,Q)
\end{equation} 
Once again theoretical masses $m_{\textrm{th}}$ and experimental masses were fitted. Results are listed in table 2. 

As expected, 
the curvature-energy constant $a_R$ is reduced by a factor 3, which results in an absolute contribution 
to the total binding energy of about $1.7$ MeV for $^{16}$O. 

For light and trans fermium nuclei we achieve 
 a significant improvement with the extended FRDMC, compared to the original FRDM. The additional 
curvature term makes the FRDMC the best model available for the description of ground state masses in the full range of the nuclear table.  

Thus, within both extended models, the FRLDMC and the FRDMC, the existence of a curvature term is supported. The coupling 
strength $\xi$, derived from fits, setting $r_0 = 1.16$ fm  results as $\xi = 7.8$ for FRLDMC and $\xi = 2.5$ for FRDMC respectively.
\begin{table}
\caption{\label{rms} root-mean-square deviations $\Delta_{\textrm{rms}}$ from AME2003 experimental data in MeV
  for different $Z$-Regions
}
\begin{tabular}{l|rrrrrr}
Z/model&8-20&20-40&40-60&60-80&80-100&$\geq$ 100\\
\hline\noalign{\smallskip}
FRLDM &1.716&0.857&0.568&0.654&0.746&1.091\\
FRLDMC&1.307&0.900&0.582&0.814&0.494&0.508\\
FRDM  &1.447&0.871&0.579&0.449&0.388&0.512\\
FRDMC &1.287&0.887&0.547&0.448&0.407&0.487\\
\end{tabular}
\end{table}
\section{Conclusion}
Based on a purely geometric interpretation of collective mass-parameters, the collective curvature 
scalar term has 
been introduced. For the geometry of the symmetric two-center shell model this term has
been derived analytically. Interpreting this term as an additional potential term 
with the explicit form $V \sim A^{-5/3}$, we have investigated the influence of this term
within the framework of two new macroscopic models: The finite range liquid drop model with curvature (FRLDMC) and 
the finite range droplet model  with curvature (FRDMC). 

Significant improvements have been found especially for light nuclei and
for trans-fermium elements. Thus, the new models allow a more precise description of nuclear ground state properties.

Therefore the collective curvature scalar as a manifestation of interaction with curved collective space plays a substantial  role in nuclear physics e.g. for strong asymmetric fission,
cluster-radioactivity or prediction of super-heavy element properties as well as other branches of physics like cosmology or star-formation.

\end{document}